\documentclass[3p, sort&compress, 12pt]{elsarticle}

\usepackage[caption=false]{subfig}

\usepackage[T1]{fontenc}
\usepackage[utf8]{inputenc}
\usepackage[english]{babel}

\usepackage{graphicx}
\usepackage{enumitem}
\usepackage{textcomp}
\usepackage{gensymb}
\usepackage{siunitx}
\usepackage{isotope}
\usepackage{rotating}
\usepackage{tabularx}

\expandafter\let\csname equation*\endcsname=\relax
\expandafter\let\csname endequation*\endcsname=\relax
\usepackage{amsmath}

\usepackage{amsfonts,amssymb,amscd}
\usepackage{xcolor, soul}
\usepackage{xspace}
\usepackage{microtype}

\usepackage{hyperref}
\usepackage{cleveref} 

\setcitestyle{square,numbers}

\sethlcolor{yellow}

\hypersetup{colorlinks, linkcolor={red!50!black}, citecolor={blue!50!black}, urlcolor={blue!80!black}}
\microtypesetup{
	protrusion=alltext-nott,
	expansion=alltext-nott,
	final
}

\graphicspath{{figures/}}

\setcitestyle{square,numbers}

\newcommand{\PWO}{\mbox{PbWO$_4$}\xspace}

\journal{Engineering Application of Artificial Intelligence}

\begin{document}

\begin{frontmatter}

\title{Deep Learning reconstruction with uncertainty estimation for $\gamma$ photon interaction in fast scintillator detectors}

\author[DES]{G.~Daniel\corref{correspondingauthor}}
\ead{geoffrey.daniel@cea.fr}
\cortext[correspondingauthor]{Corresponding author}
\author[BIOMAPS,DES] {M.B. Yahiaoui}
\author[BIOMAPS]{C. Comtat}
\author[BIOMAPS]{S.~Jan}
\author[BIOMAPS]{O.~Kochebina}
\author[DES]{J-M. Martinez}
\author[BIOMAPS,DES]{V. Sergeyeva}
\author[IRFU,BIOMAPS]{V. Sharyy}
\author[IRFU]{C-H.~Sung}
\author[IRFU,BIOMAPS]{D.~Yvon}

\address[DES]{Université Paris-Saclay, CEA, Service de Génie Logiciel pour la Simulation, 91191, Gif-sur-Yvette, France}
\address[BIOMAPS]{Université Paris-Saclay, Inserm, CNRS, CEA, Laboratoire d’Imagerie Biomédicale Multimodale (BioMaps), 91401, Orsay, France}
\address[IRFU]{Université Paris-Saclay, CEA, IRFU, Département de Physique des Particules, 91191, Gif-sur-Yvette, France}


\begin{abstract}

This article presents a physics-informed deep learning method for the quantitative estimation of the spatial coordinates of gamma interactions within a monolithic scintillator, with a focus on Positron Emission Tomography (PET) imaging. A Density Neural Network approach is designed to estimate the 2-dimensional gamma photon interaction coordinates in a fast lead tungstate (\PWO) monolithic scintillator detector. We introduce a custom loss function to estimate the inherent uncertainties associated with the reconstruction process and to incorporate the physical constraints of the detector.

This unique combination allows for more robust and reliable position estimations and the obtained results demonstrate the effectiveness of the proposed approach and highlights the significant benefits of the uncertainties estimation. We discuss its potential impact on improving PET imaging quality and show how the results can be used to improve the exploitation of the model, to bring benefits to the application and how to evaluate the validity of the given prediction and the associated uncertainties. Importantly, our proposed methodology extends beyond this specific use case, as it can be generalized to other applications beyond PET imaging.

\end{abstract}


\end{frontmatter}

\section{Introduction}

Gamma photon detection is used in numerous industrial, medical and security applications. 
It is often based on scintillating crystals coupled with a light collection and readout system. 
The scintillator is either pixelated, consisting on an array of small individual crystals, or continuous, made of a large monolithic block.
The choice between both technologies is often based on a trade-off between sensitivity and resolution performances.  
For pixelated detectors, the spatial localization of the detection is provided by the crystal impacted by the gamma photon, whereas for continuous detectors,
designated algorithms shall be used to derive spatial information.
Several algorithms have been proposed, based either on prior knowledge of the physics of detection or on a machine learning approach~\cite{kawula_sub-millimeter_2021}.
The main objective of this work is the development of a physics informed deep learning method for a quantitative estimation of the spatial coordinates of the gamma interaction within a monolithic scintillator, including uncertainties.
The application framework is nuclear medicine imaging, more specifically the detection of 511~keV gamma photons in Positron Emission Tomography (PET).
\\
PET imaging is a powerful {\it in vivo} functional imaging modality  mainly used  in oncology, neurology and cardiology.
It is based on the administration to the patient of a biomarker labeled with a radionuclide that decays through positron emission, 
followed by the detection in  coincidence, outside the body, of pairs of 511~keV gamma photons resulting from the
annihilation of the emitted positrons with electrons of surrounded media.
The data acquisition process is followed by the tomographic reconstruction of a three-dimensional image of the biomarker distribution within the body.
The quality of the reconstructed image highly depends on the performances of the gamma photon detectors in terms of sensitivity, spatial resolution, and temporal resolution.
The detector spatial resolution has a direct impact on the contrast recovery of small structures in the image.
A higher detection efficiency translates into a higher number of detected coincidences, resulting in a better signal-to-noise ratio (SNR) in the PET image.
The improvement of the coincidence resolving time (CRT), characterizing the capability of a pair of detectors to resolve the difference 
between the times of interaction of the two 511 keV gamma photons detected in coincidence,
also helps for increasing the SNR in the image \cite{Schaart_2021}.
This principle is referred to as time-of-flight (ToF) PET.
\\
State of the art clinical PET systems are based on pixelated detectors made of LSO (lutetium
oxyorthosilicate) or LYSO (lutetium-yttrium oxyorthosilicate) crystals, 
with a pixel pitch between 3.5 and 5~mm.
These PET systems have, at best, a CRT of 210~ps~\cite{vanSluis2019Jan}, 
an intrinsic spatial resolution in the reconstructed image of 3.5~mm~\cite{vanSluis2019Jan} (about 3~mm at the detector level), 
and an absolute sensitivity of $\sim$~20~counts/sec/kBq (i.e. $\sim$~2\%)~\cite{Grant2016}. 
Major efforts are being made worldwide in nuclear instrumentation research groups to improve these parameters, 
in particular to decrease the CRT below 100~ps, ideally down to 10~ps~\cite{lecoq_roadmap_2020}. 
There is a trend toward the development of PET detectors based on monolithic crystals for their higher sensitivity,
since there are no intercrystal gaps~\cite{gonzalez-montoro_evolution_2021}.
Based on the scintillation light distribution readout, dedicated neural networks have recently been proposed to provide
a single 3-dimensional gamma photon interaction position within the monolithic crystal (see, for example~\cite{belov_resolution_2023,freire_performance_2022,carra_neural_2022,jaliparthi_deep_2021,kawula_sub-millimeter_2021}). 
All these studies demonstrate the benefit of using a machine learning approach for an accurate position estimation.
\\
In this study, we propose to address the question of the reconstruction of the 2-dimensional gamma photon interaction coordinates for a fast lead tungstate (\PWO) monolithic scintillator detector \cite{Yvon2020}, that is not straightforward due to the specificity of our acquisition system which is not a conventional pixelated photodetector. Our system consists indeed in the use of a micro-channel plate photomultiplier tube (MCP-PMT) which necessitates a dedicated process to reconstruct gamma photon interaction parameters from the acquired signal, as we describe in this paper. Moreover, we aim not only to perform the reconstruction but to associate uncertainties on the reconstruction. We will exploit and validate these uncertainties to assess the reliability of the network prediction. This approach is new in this field and the methodology we propose can be extended to other applications on signal processing or sensor data analysis.
In the future, these spatial coordinates uncertainties could be used during the tomographic reconstruction and potentially improve the quality of the PET image. 
\\
The paper is organized as follows. 
In section~\ref{sec:Materials}, we describe the gamma photon detector based on \PWO, the Monte Carlo simulation used to generate the training and testing datasets, and the preprocessing of the detector raw data. 
Section~\ref{sec:Methodology} focuses on the methodology of the deep learning approach used for this study,
including the description of a baseline method for performance comparisons and the definition and the architecture of the selected Density Neural Network. 
Results are presented in section~\ref{sec:Results}. These results and the methodology are discussed in section~\ref{sec:Discussion}.
\\
\\
 \begin{figure}
        \includegraphics[width=.54\textwidth]{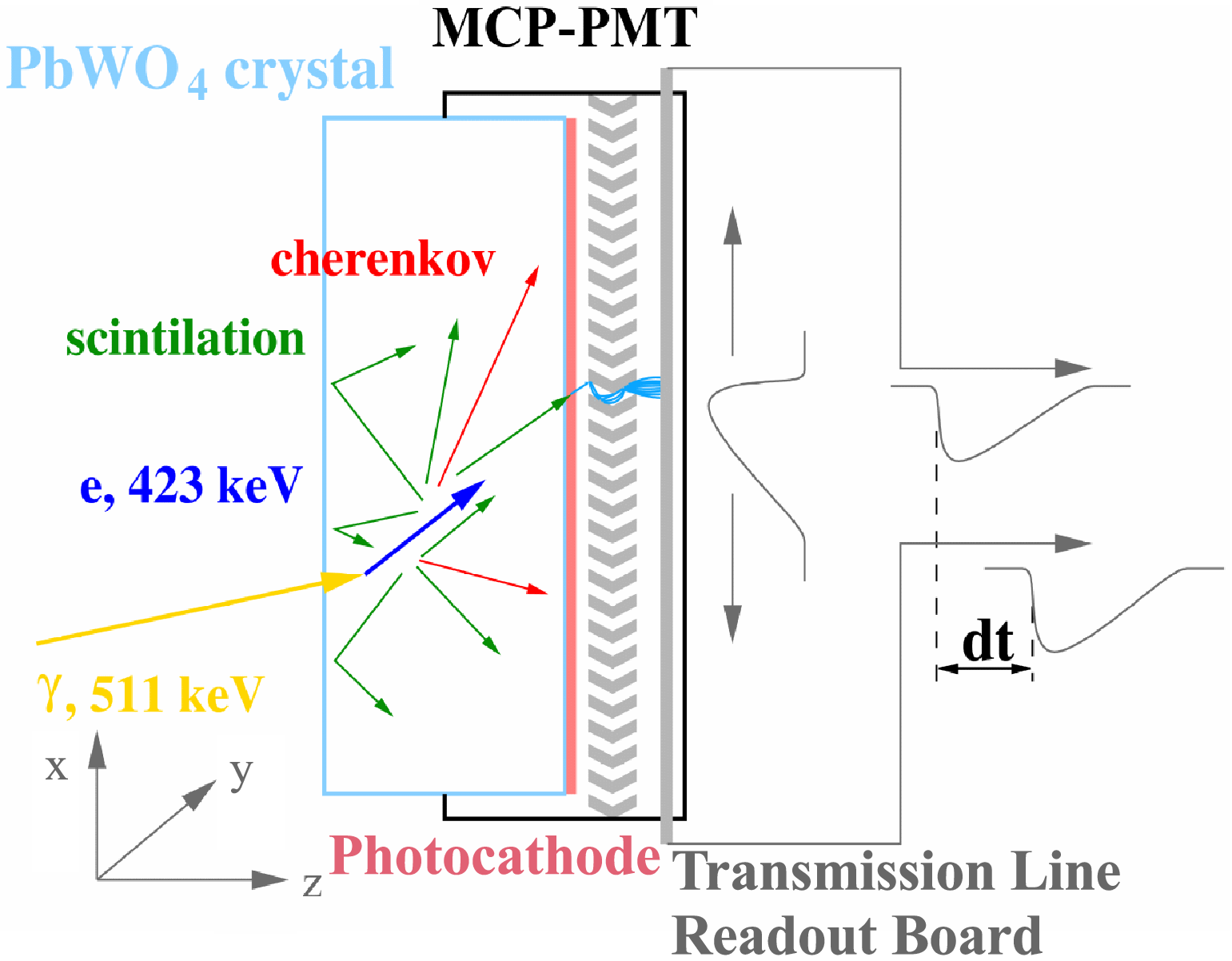}
        \includegraphics[width=.44\textwidth]{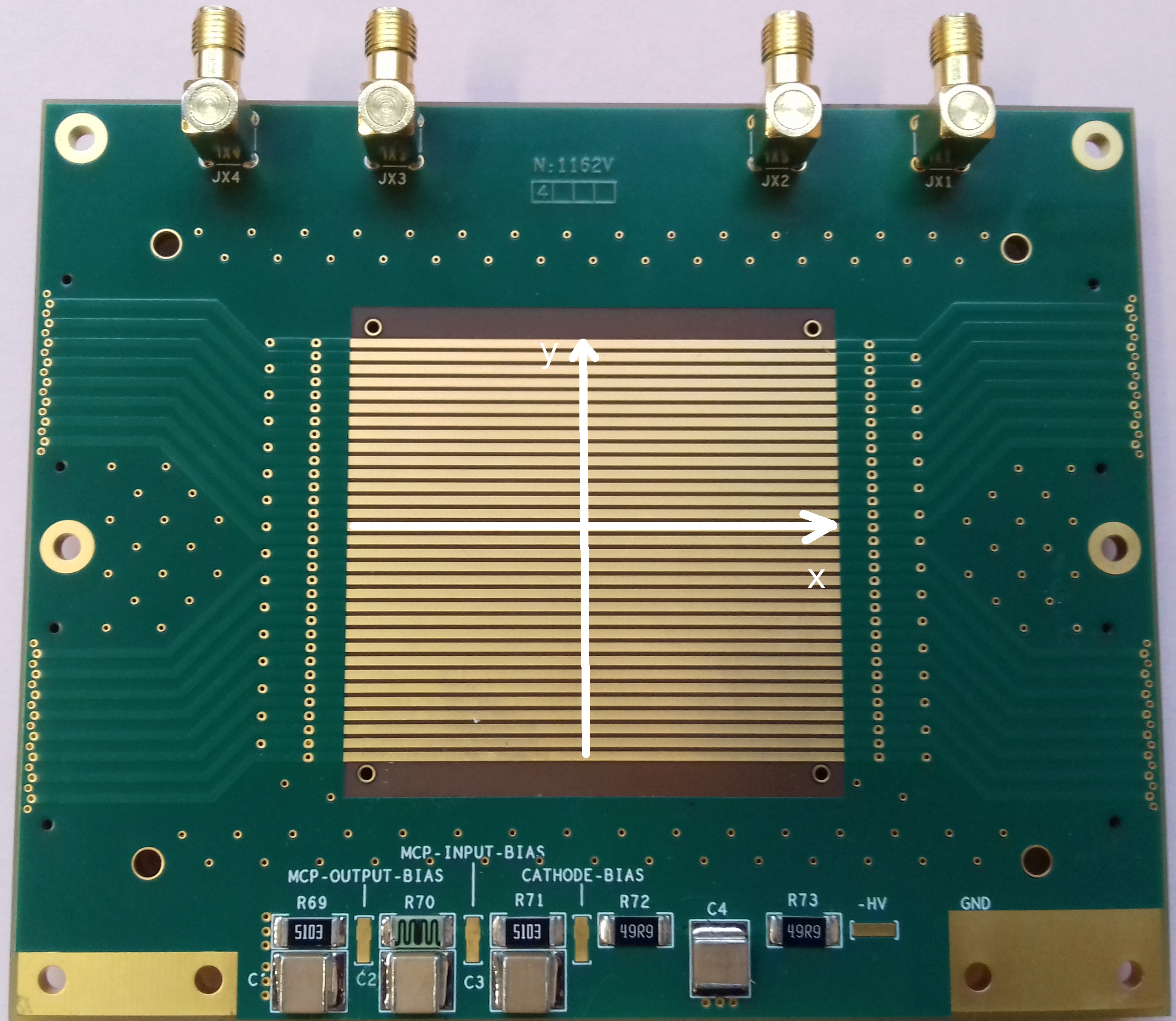}
        \caption{
        \textbf{Left:} Schematic diagram of the ClearMind detection module. A 511 keV gamma-ray interaction in the crystal produces scintillation and Cherenkov photons that are converted by the photocathode to photoelectrons. These photoelectrons are then multiplied by the MCP-PMT and induce signals on the transmission lines (TLs). Signals from the left and right ends of each TL are amplified by 40 dB amplifiers and digitized by a SAMPIC module.
        \textbf{Right:} Transmission lines Printed Circuit Board (PCB). The axis $x$ and $y$ corresponds to the coordinate system that we use to locate the interaction position.}
        \label{fig:DetSchematic}
\end{figure}

\section{Materials}
\label{sec:Materials}    
    \subsection{Detector description}
        
\begin{figure}[h!]
        \centering
        \includegraphics[width=0.49\textwidth]{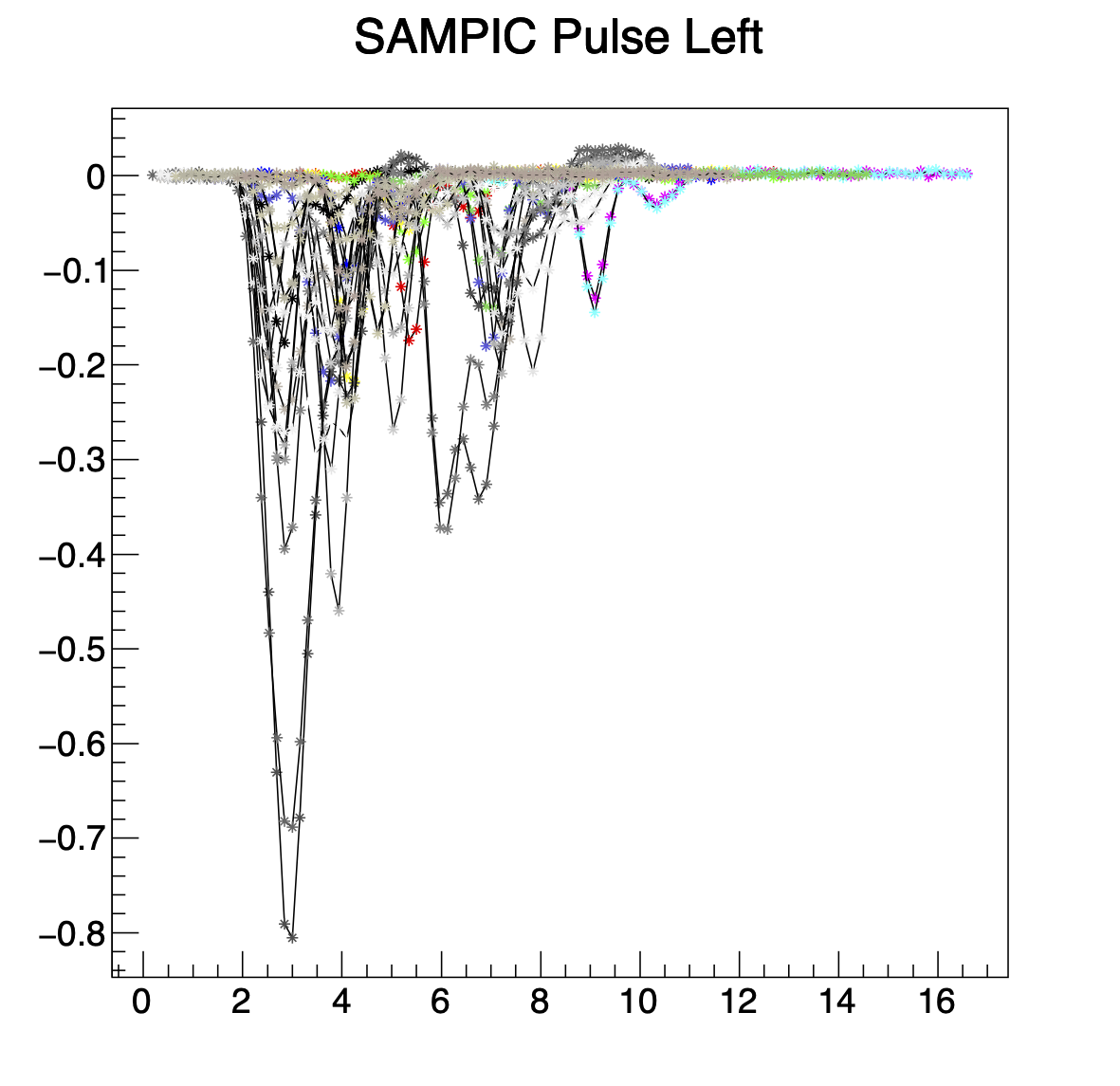}
        \caption{Set of pulses as registered by the SAMPIC waveshape recorder for a 511 keV energy deposit. For clarity purpose, only the pulses registered on one side of the transmission lines are shown (half of the set).}
        \label{fig:EvtPulseSet}
\end{figure}

The ClearMind gamma detector (Figure \ref{fig:DetSchematic}) is composed of a MCP-PMT sealed by a monolithic \PWO crystal, acting both as the gamma conversion crystal and as the optical window of the MCP-PMT.  
A high quantum efficiency photoelectric layer is deposited on its inner face. 
The direct deposition of a photocathode with a refraction index superior to the refraction index of the \PWO crystal allows us to avoid total reflection at the crystal/photocathode interface, thus maximizing the photon collection efficiency of the module \cite{Yvon2020}. 
The use of this ”scintronic” crystal as an entrance window of a MCP-PMT makes it possible to optimize the time resolution thanks to the excellent electron transit time spread ($\sim$60~ps FWHM) to the detection anodes provided by this type of photodetector. \newline
The \PWO crystal, homogeneously doped, has a surface of $59~\mathrm{mm} \times 59~\mathrm{mm}$\, a thickness of 5~mm, and is provided by CRYTUR \cite{CryturComp}. 
The photocathode deposit and the integration of the device into a MCP-PMT structure is handled by the PHOTEK company, based on its MAPMT-253 design \cite{MAPMT253}.
We developed a signal readout system for this device using 32 transmission lines as shown in Fig.~\ref{fig:DetSchematic}, \cite{FOLLIN_2022}. 
The signals are read out at both ends of the transmission lines, amplified and then sampled by a SAMPIC WaveShape Recorder \cite{Breton_2020}.\newline
Typically a 511~keV energy deposit in the crystal produces 185~optical photons mostly isotropically. 
Out of these, $\sim$~20\% are collected by the $53~\mathrm{mm} \times 53~\mathrm{mm}$ photocathode and generate detected photoelectrons that are collected and amplified by the MCP-PMT photodetector. \newline
Many processes in the signal formation involves random features. 
For example, the photon direction and time of production, the photoelectron production probability, the gain of the Micro Channel Plate used for electron multiplication, the time transit of the electron propagation through the MCP-PMT, the noise of readout amplifiers are best described using parametrized random variables. 
These are necessary to describe the pulses shapes produced by a single photoelectron (SPE). \newline
Each SPE induces a signal on typically three readout lines. 
Thus the typical 30 SPE signals pileup at the output of the transmission lines (Figure \ref{fig:EvtPulseSet}), and build the event signal registered by the SAMPIC module.

    \subsection{Simulation}
   In order to create a sufficiently large and unbiased database to train and test
    our reconstruction algorithms, we developed a detailed simulation of the ClearMind detector \cite{Sung:2022dij}. The knowledge of the hidden physical processes available in Monte Carlo simulations is necessary to provide the ground truth (target) for the training of the Machine Learning models. It is also very convenient to assess the intrinsic performances of our algorithms. 
    
    This simulation is based on the  Gate v9.0 \cite{Jan2004Sep,Sarrut2021May} /
    Geant4 v7.0 \cite{agostinelli2003, Allison2006Feb, Allison2016Nov}
    software allowing to simulate in full details the interaction of the particle with matter and optical photons generation and tracking. Furthermore, we have developed specialized software to simulate the photodetector, analog, and digital electronic components. Necessary parameters have been extracted from dedicated measurements \cite{Follin_2021, FOLLIN_2022}.
    This simulation includes the following main parts of the detector response.
    \begin{enumerate}
        \item The gamma interaction in the crystal accounts for three processes: photoelectric conversion, Compton scattering and Rayleigh diffusion. The two first processes produce relativistic electron that emits visible
              photons through two mechanisms: Cherenkov radiation and scintillation
              ($\sim$20 and $\sim$165 photons for 511 keV $\gamma$-quanta respectively). 

        \item Each optical photon is propagated individually by the simulation program. During the propagation all main physical effects are taken into account: 
            photon absorption inside the \PWO crystal,
              reflection or absorption on the crystal borders for the different types of the crystal surface
              (polished, ground, absorbing), escape of photons from the crystal into the air. 

        \item Photocathode simulation includes the Fresnel reflection of visible photons at photocathode boundaries, absorption of photons by the photocathode and extraction of generated photoelectrons as a function of the photon wavelength. 
        As a result we compute, assuming a photocathode of nominal efficiency, that we produce in average 30 photoelectrons
              for a 511 keV $\gamma$-ray photoelectric conversion in the crystal and 
              75\% of events contain at least one Cherenkov photon converted into a photoelectron. 

        \item We then simulate the propagation and the multiplication of individual photoelectrons generated by the photocathode in the MCP-PMT and parametrize the main PMT response features: time response, PMT gain and gain fluctuation, signal sharing between different output anodes.

        \item Finally, we simulate the signal readout through the transmission lines with realistic signal shapes, taking into account the possible overlay of several photoelectrons, electronics noise and digitization sampling. 
              
    \end{enumerate}

Most of the  simulation parameters are adjusted to the results obtained by the characterization of the first prototype using
pulsed laser in the single-photon regime. More details about the simulation could be found elsewhere \cite{Sung2022,Sung:2022dij}.

\subsection{Waveform preprocessing and input data shaping}
\label{sec:WaveProcess}
Depending on the energy deposited in the crystal, the SAMPIC module records from 2 to 64 pulse shapes, typically about 30 for a deposited energy of 511 keV. 
This corresponds to about 10 kBytes of raw data per event. 
This volume of data is to be compared with the volume of data to be reconstructed, the properties of the gamma interaction in the crystal: 
3D position, time, deposited energy, interaction multiplicity, typically 25 bytes per event. 
The acquired data are therefore redundant, but also intricate. 
The parameters to be reconstructed are encoded in a complex way and are mixed in the pulse shapes.

Thus reconstructing the properties of the gamma interaction in the crystal is a complex task.

We first use our knowledge of the physics of the detector to calculate on the raw data a set of statistical variables, so called "parameter observables" highly correlated to the parameters of the gamma-ray interaction to be reconstructed, as well as a second set of "bias observables" which monitor the known undesirable instrumental effects (saturation of the acquisition electronics, edge effects, etc...). 
The data volume of the observables is 100 bytes per event. These observables are then used as inputs of shallow, fully connected neural networks, whose training takes only a few minutes. The development cycle is fast, at the cost of a loss of information that is difficult to anticipate. \newline

In the  following paragraph, we explain the parameter observables developed to correlate to the features targeted by the neural networks presented in this paper: 
the gamma interaction position $x_\mathrm{pos}$ (\emph{position along the transmission lines}) and $y_\mathrm{pos}$ (\emph{perpendicular to the transmission lines}).

\begin{figure}[ht]
        \centering
        \includegraphics[width=0.9\textwidth]{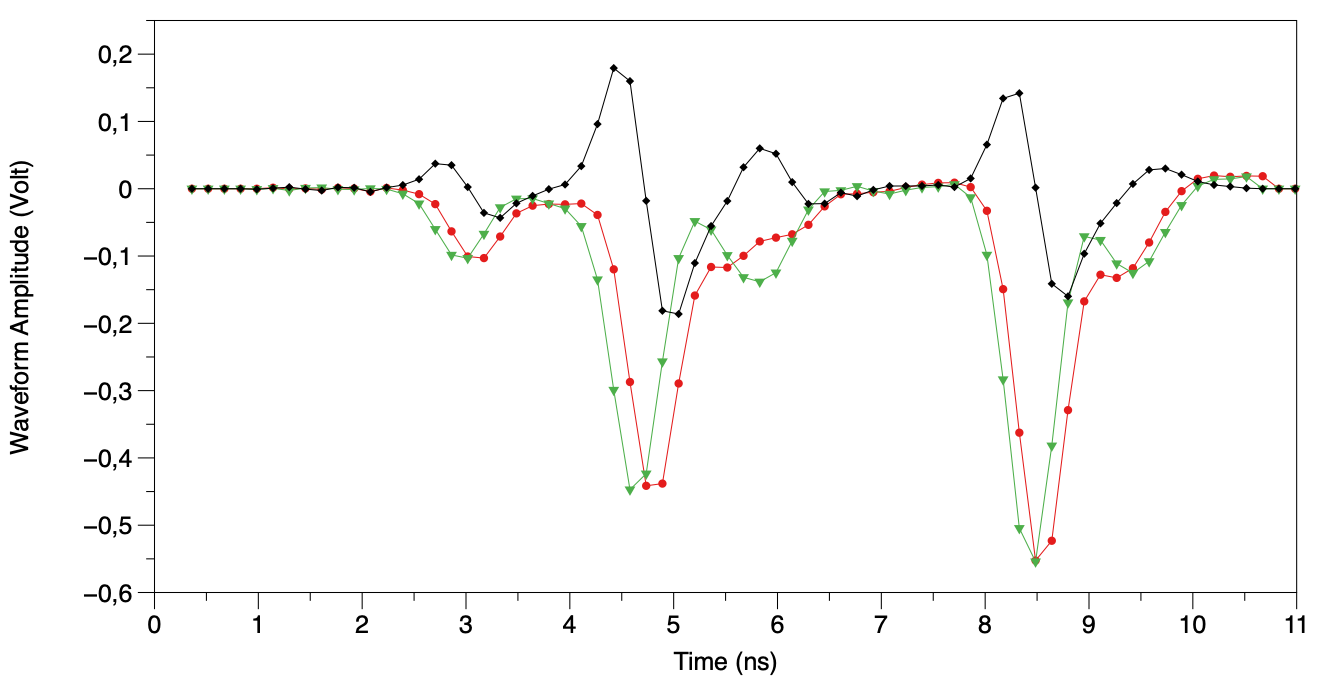}
        \caption{Waveforms registered on one trigered line $l$. Red and green lines are the left $F_{l,\mathrm{Left}}(t_{j})$ and right $F_{l,\mathrm{Right}}(t_{j})$, time shifted, registered pulses shapes. Black line shows the time difference waveform $F_{l,\mathrm{Left}}(t_{j}) - F_{l,\mathrm{Right}}(t_{j})$. We identify on this line three pulses clusters at 4.6 ns, 6 ns and 8.5 ns. For each of them, the time difference curve shows a bipolar shape, correlated to the position of each photoelectron charge induction along the readout line.}
        \label{fig:LineProcessing}
\end{figure}

First, for each transmission line $l$, that triggered the SAMPIC acquisition,  the digitized pulse shapes are acquired: $F_{l,\mathrm{Left}}(t_{j})$ and $F_{l,\mathrm{Right}}(t_{j})$ at times $t_{j}$ for the Left and Right transmission line, where the index $j$ corresponds to the sampling time. Examples of acquired pulse shapes are shown in Figure \ref{fig:LineProcessing}.

We first calculate the electric charges collected (integral of the pulse shape current over time) at both the end of the 32 transmission lines $C_{l}$.\newline

We compute two interaction observables expected to be correlated to $y_\mathrm{pos}$ :
        \begin{itemize}
            \item 
            We select the line with the largest collected charge and its two neighbors. We fit a parabola on these three values. The position of the maximum of this parabola is the first observable.
            \item
            The second observable is the median of the distribution of line numbers weighted by the charge collected on these lines : $\mathrm{Med}_{L} = \mathrm{median}(l, C_{l})$.
            It is common that one line carries a large fraction of the total charge collected over the 32 readout lines registered in an event. In order to extract as much information as possible from the surrounding line charge values, we developed an "upgraded" median algorithm documented in section \ref{sec:MedianAlgo}. This is the "median" algorithm we will use for all the observables calculations.
        \end{itemize}

Quantifying observables expected to correlate to $x_\mathrm{pos}$ is more complex.
 It turns out that on some lines, we can identify few pulses separated in time in the recorded pulse shapes. We denote each identified pulse by the index $p$. When this happens, we quantify the pulse detection times ($T_{l,p,\mathrm{Left}}$ and $T_{l,p,\mathrm{Right}}$) and the total charge $C_{l,p}$ of each of these pulses. 
The time difference $\Delta t_{l,p} = T_{l,p,\mathrm{Left}} - T_{l,p,\mathrm{Right}}$ is correlated to the position of the pulse induced by this photoelectron along this line. We then calculate the three following observables:
\begin{itemize}
    \item 
    the \emph{median} of the 
    $\mathrm{Med}_{\Delta t} = \mathrm{median}(\Delta t_{l,p}, C_{l,p})$ distribution (all lines and pulses), weighted by the integrated charge of the each pulses. The details of the algorithm for the weighted median are given in Section \ref{sec:MedianAlgo};
    \item
    the \emph{mean} of the same distribution
     $\mathrm{Mean}_{\Delta t} = \mathrm{mean}(\Delta t_{l,p}, C_{l,p})$;
    \item
    $\Delta t_{\mathrm{lmax}}$, the time difference $\Delta t$ of the largest pulse of the line (greatest charge collected).
 \end{itemize}  
 
We also used the shape of the digitized pulses $F_{l,p,\mathrm{Left}}(t_j)$ and $F_{l,p,\mathrm{Right}}(t_j)$.
The $\mathrm{Diff}_{l,p}(t_j) = F_{l,p,\mathrm{Left}}(t_j) - F_{l,p,\mathrm{Right}}(t_j)$ show bipolar shapes depending on the position of the pulse injection along the line. We integrated the $\mathbf{F}$irst and $\mathbf{S}$econd components of the bipolar shape $\mathrm{Int}_{l,p,F}$ and $\mathrm{Int}_{l,p,S}$. We then compute the $\mathrm{Bipol}_{l,p} = (\mathrm{Int}_{l,p,F} - \mathrm{Int}_{l,p,S})/C_{l,p}$, for each pulse. 
The saved observables are then:
\begin{itemize}
    \item 
    the \emph{weighted median} of the distribution $\mathrm{Med}_{\mathrm{Bipol}}=\mathrm{median}({\mathrm{Bipol}}_{l,p})$ (see section \ref{sec:MedianAlgo} for the details)
    \item 
    the \emph{mean} of the distribution $\mathrm{Mean}_{\mathrm{Bipol}}=\mathrm{mean}({\mathrm{Bipol}}_{l,p})$.
\end{itemize}

Along with these observables, we also compute parameters and bias observables relevant for reconstruction the other properties gamma interactions in the crystal (see section \ref{sec:Prepro}). Some of them are also expected to be relevant for the uncertainty estimation.
These \textbf{23 observables are the inputs of the following neural networks}.\newline

An alternative approach would be to use the full waveforms' data as inputs for convolutional deep neural networks. However, preliminary tests showed that the complexity of the data required to implement deeper network structures, involving training times of days on our GPU card, thus with slower development cycles. We are considering this option for future works.

\section{Methodology}
\label{sec:Methodology}
    \subsection{Baseline method} 
    The simplest approach to gamma-interaction reconstruction is using the fact that the highest density of the detected photons corresponds to the coordinates of the interaction point.
This suggests to use the transmission line with the maximum detected charge as the reference line.
To reconstruct the coordinate across lines, $y_R$, we calculate the weighted average of center-of-line for this and two neighboring lines:
\begin{equation} \label{eq:yr}
  y_R=\frac{\sum_{l=i-1}^{i+1} y_lC_l}{\sum_{l=i-1}^{i+1} C_l} \;,
\end{equation}
where $y_l$ is the y-coordinate of the line center,
$C_l$ is the charge of line $l$
(only the negative signal part is used for the charge calculation),
$i$ is the reference line number.

The coordinate along the lines, $x_R$ is reconstructed as
\begin{equation} \label{eq:xr}
  x_R = \frac{(t_\mathrm{Right} - t_\mathrm{Left})}{2} \times v_\mathrm{signal} \;,
\end{equation}
where $t_\mathrm{Right}$ and $t_\mathrm{Left}$ are the time measured at the right and left ends of
line $i$ respectively and $v_\mathrm{signal}$ is the signal propagation speed,
assumed to be  35\% of speed of light in the simulation.

This approach, based only on expert knwoledge of the physical processes is used as a reference for comparison with the performance of our approach based on neural networks.

    \subsection{ML approach} 
    This section describes our approach based on a supervised Machine Learning algorithm, that takes as input the 23 preprocessed variables described in sections \ref{sec:WaveProcess} and \ref{sec:Prepro} in order to predict the interaction position. We use a Neural Network model that is able to provide both the prediction and an estimation of an uncertainty associated to this prediction thanks to the paradigm of Density Neural Networks.On the contrary to the baseline method, the use of Neural Network is mainly based on the exploitation of simulated data to build a model that link the observable quantities to the reconstruction of the gamma interaction position. Moreover, the baseline method is not able to provide uncertainties associated to the reconstruction, which is a key difference with our approach.

        \subsubsection{Density Neural Network}
   
    The conventional approach to perform a regression through a Neural Network is to define a loss function that measures the discrepancy between the predictions of the neural network and the expected values. With classical notations, let $\mathcal{D}=\{(x_i,y_i),i=1,2,\ldots,N\}$ be the database and $\mu_{\theta}(x)$ the prediction of the expected value of the neural network parameterized by $\theta$. By noting $||\bullet||$ the Euclidean norm, a usual loss function is the Mean Square Error :
        \begin{equation}
        \label{eq:MSE}
            \textrm{MSE}(\theta) = \frac{1}{N}\sum_{i=1}^N||y_i - \mu_{\theta}(x_i)||^2.
        \end{equation}
    The learning procedure consists in finding parameters $\theta^*$, among the space of possible parameters $\Theta$, that minimizes the Mean Squared Error. It corresponds to solving the following problem:
        \begin{equation}
        \label{eq:minprobMSE}
            \theta^* = \underset{\theta \in \Theta}{\textrm{argmin}} \; \textrm{MSE}(\theta). 
        \end{equation}

    The Density Neural Network approach proposes to make the assumption that the expected outputs $y_i$ are a realization of a probability distribution such as the normal distribution $y_i \sim \mathcal{N}(\mu_{\theta}(x_i), \sigma_{\theta}^2(x_i))$ or a mixture of normal distributions \cite{bishop1994}. However, this method can be adapted to other distributions and we propose to use a distribution that considers some specific physical constraints of our application.
    
    The main physical constraint on the interaction position is that this interaction must be located in the limits of the detector. Therefore, we decide to use a truncated normal distribution in order to insure this constraint. 
    
    Beforehand in order to avoid confusion with $(x,y)$ coordinates of the position of the interaction in the detector, we will note $s$ the inputs data from the preprocessing describes in Section \ref{sec:WaveProcess}. So formally, our neural network is now expected to output four values for each inputs $s_i$, the two means $\mu_{\theta,x}(\mathrm{s_i}), \mu_{\theta,y}(\mathrm{s_i})$ and the two scale parameters $\sigma_{\theta,x}^2(\mathrm{s_i}), \sigma_{\theta,y}^2(\mathrm{s_i})$ of the assumed truncated Gaussian law. We have assumed independent normal laws on $x_i$ and $y_i$. In this case, the hypothesis that each output $x_i,y_i$ follows a truncated normal distribution in $[a,b]$ leads to the two following distributions relative to the two coordinates with $z=x$ or $z=y$:
        \begin{eqnarray}
            \label{eq:truncgauss}
            p(z_i|\mu_{\theta,z}(\boldsymbol{\mathrm{s_i}}),\sigma_{\theta,z}^2(\boldsymbol{\mathrm{s_i}}) )  & = &  \underbrace{\frac{1}{\sqrt{2\pi\sigma_{\theta,z}^2(\boldsymbol{\mathrm{s_i}})}} \exp\left(-\frac{(z_i -\mu_{\theta,z}(\boldsymbol{\mathrm{s_i}}))^2}{2\sigma_{\theta,z}^2(\boldsymbol{\mathrm{s_i}})}\right)}_{\textrm{Classical Gaussian Likelihood  }} \nonumber \\
             & \times &  \underbrace{\frac{1}{\Phi\left(\frac{b - \mu_{\theta,z}(\boldsymbol{\mathrm{s_i}})}{\sigma_{\theta,z}(\boldsymbol{\mathrm{s_i}})}\right) - \Phi\left(\frac{a - \mu_{\theta,z}(\boldsymbol{\mathrm{s_i}})}{\sigma_{\theta,z} (\boldsymbol{\mathrm{s_i}})}\right)}}_{\textrm{With truncation in  }[a,b]}
        \end{eqnarray}
    where $\Phi$ is the Cumulative Distribution Function of the standard normal distribution, $a$ and $b$ are the truncation boundary of the truncated normal distribution. In our application, $a$ and $b$ represent the limits of the detector: $a = -30 \; \textrm{mm}$ and $b = 30 \; \textrm{mm}$. In our case, the condition $a \leq (x_i,y_i) \leq b$ is assumed to be always verified : an interaction occurs inside the detector; otherwise it is a simulation error and we should not consider it. So, by using equation \ref{eq:truncgauss}, we derive the likelihood of the parameters on the whole dataset:
        \begin{equation}
            \label{eq:likelihoodtruncgauss}
            L(\theta) = \prod_i \prod_{z \in \{x,y \}} p(z_i|\mu_{\theta,z}(\boldsymbol{\mathrm{s_i}}),\sigma_{\theta,z}^2(\boldsymbol{\mathrm{s_i}}) )
        \end{equation}
    Maximizing this likelihood is equivalent to minimizing the negative log-likelihood, that is easier to handle, which leads to the following loss function:
        \begin{eqnarray}
            \label{eq:loglikelihoodtruncgauss}
            l(\theta)  & = & -\log(L(\theta)) \nonumber \\
            & = & \sum_i \sum_{z \in \{x,y\}} \log\left(\Phi\left(\frac{b - \mu_{\theta,z}(\boldsymbol{\mathrm{s_i}})}{\sigma_{\theta,z}(\boldsymbol{\mathrm{s_i}})}\right) - \Phi\left(\frac{a - \mu_{\theta,z}(\boldsymbol{\mathrm{s_i}})}{\sigma _{\theta,z}(\boldsymbol{\mathrm{s_i}})}\right)\right) \nonumber \\
            & + & \frac{1}{2} \log(2\pi\sigma_{\theta}^2(\boldsymbol{\mathrm{s_i}})) + \frac{(z_i -\mu_{\theta,z}(\boldsymbol{\mathrm{s_i}}))^2}{\sigma_{\theta,z}^2(\boldsymbol{\mathrm{s_i}})}
        \end{eqnarray}

        We want to highlight that the conventional problem with $\textrm{MSE}$ defined in equation \ref{eq:MSE} is a special case of equation \ref{eq:loglikelihoodtruncgauss} and $a = -\infty$, $b = +\infty$ (no truncature) and where the scale parameters $\sigma_{\theta,x}^2(\boldsymbol{\mathrm{s_i}})$ and $\sigma_{\theta,y}^2(\boldsymbol{\mathrm{s_i}}) $ are assumed to be constant (homoscedasticity hypothesis). Finally, the optimization problem consists now in finding the parameters $\theta^*$ that minimize this new loss function:

        \begin{equation}
        \label{eq:minprobloss}
            \theta^* = \underset{\theta \in \Theta}{\textrm{argmin}} \; l(\theta)
        \end{equation}

        We can solve this problem by using a conventional gradient descent optimization in order to train the neural network and find optimal parameters.

        This approach brings two main components:
        \begin{itemize}

            \item 
            the use of a Density Neural Network allows the estimation of an uncertainty associated to the network output through the prediction of the terms $\sigma^2_{\theta,x}(\boldsymbol{\mathrm{s_i}})$ and $\sigma^2_{\theta,y}(\boldsymbol{\mathrm{s_i}}) $. These variances are specific to each example and represent a higher or lower uncertainty on the predicted positions $\mu_{\theta,x}(\boldsymbol{\mathrm{s_i}})$ or $\mu_{\theta,y}(\boldsymbol{\mathrm{s_i}})$ according to the inputs $\boldsymbol{\mathrm{s_i}}$ of the neural network. For instance, in our application, we can expect an interaction closer to the edge of the detector is less precisely located than an interaction at the center of the detector. We highlight that this uncertainty is self-estimated by the neural network and consequently must be empirically validated, we bring elements for this validation in section \ref{sec:uncertaintyperf}. This uncertainty can be seen as aleatoric uncertainty \cite{DBLP:journals/corr/KendallG17} due to the randomness of the phenomenons involved during the detection and measurements (Cherenkov and scintillation production, loss of optical photons, randomness in the pulse shape creation...).
            
            \item
            the use of a truncated normal hypothesis allows us to consider the physical constraints. In section \ref{sec:Results}, we show that this property is essential to obtain better performances for interaction close to the edges of the detector, compared to the use of the classical MSE approach.
        \end{itemize}

        \subsubsection{Architecture and training}

        We develop two neural networks: one neural network associated to the conventional Mean Squared Error loss function and one neural network associated to our custom loss function with the truncated normal hypothesis. For both architectures, all of the hidden layers are similar: we use six hidden fully-connected layers with 256 neurons each and a tanh activation function.
        The difference is the ouptut layer:
        \begin{itemize}

            \item 
            the output layer for the neural network associated to the MSE loss has two neurons (one for each coordinate), with a scaled tanh function: $\frac{b-a}{2}\tanh$ in order to insure that the prediction is located inside the detector.
            
            \item
            the output layer for the neural network associated to the truncated normal loss function has four output neurons. The first ones output $\mu_{\theta,x}(\boldsymbol{\mathrm{s_i}})$ and $\mu_{\theta,y}(\boldsymbol{\mathrm{s_i}})$ and the activation function is the previous scaled tanh function. The second ones output $\sigma_{\theta,x}(\boldsymbol{\mathrm{s_i}})$ and $\sigma_{\theta,x}(\boldsymbol{\mathrm{s_i}})$. The activation function is $\textrm{softplus}(\alpha) + \epsilon$ where $\alpha$ is the output before applying the activation function. $\epsilon$ is a small constant that ensures that $\sigma_{\theta,x}(\boldsymbol{\mathrm{s_i}})$ and $\sigma_{\theta,y}(\boldsymbol{\mathrm{s_i}})$ are not too close to 0 and avoids divergence issues. In our case, we choose $\epsilon = 10^{-6} \; \textrm{mm}$.
        \end{itemize}

        The neural networks and their trainings are computed by using the Tensorflow \cite{tensorflow2015-whitepaper} library. The optimizer is the Adam algorithm. The training is stopped by an early stopping procedure that monitors the validation loss; we use randomly 20 \% of our data as validation data. The code has been executed on a NVIDIA RTX Quadro 5000. The training time is a dozen of minutes.

            \subsubsection{Generated datasets}

We used three datasets to train and test our supervised Machine learning algorithms. All of them use the detector modeling explained in section \ref{sec:Materials}, corresponding to a PbWO4 scintallation crystal of 59 mm $\times$ 59 mm. The inputs are the 23 observables described in sections \ref{sec:WaveProcess} and \ref{sec:Prepro} from the 32 transmission lines and the outputs are both coordinate X and Y of the gamma interaction, which are saved thanks to the Geant4 simulation.

\begin{enumerate}
  \item For the training set, we simulate a Gamma photon source  shaped in a 6 cm cube. 
  The energy spectrum has been adjusted to increase the probability of high energy deposits in the \PWO and thus generate an approximately flat deposited energy spectrum. Thus we generate 7 times more 1.2 MeV than 300 keV photons. The gamma rays impinge the \PWO crystal perpendicularly and uniformly over the entire surface of the optical window.
  This training batch contains 450 000 events.

  \item The first test dataset simulates an grid of $9 \times 9$  511-keV gamma ray point sources, regularly spaced by 7 mm. Again the gamma rays impinge the \PWO crystal perpendicularly to the surface of the optical window.
  This test dataset contains 300 000 events.

  \item The second test dataset simulates a 6 cm cube-shaped Gamma photon source, mono-energetic at 511 keV. The gamma rays impinge the \PWO crystal perpendicularly and uniformly on the whole surface of the optical window.
  This test dataset contains 600 000 events. We design this test dataset with a high number of examples in order to conduct an accurate analysis of the performances of the neural network considering the uncertainty estimation.
\end{enumerate}

    \begin{figure}[h!]
        \centering
        \includegraphics[width=0.5\textwidth]{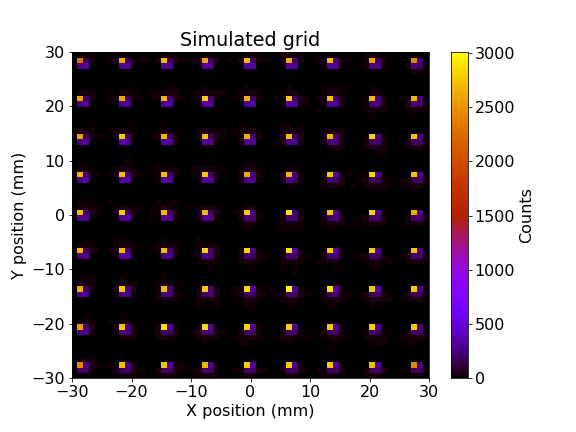}
        \caption{Simulated sources - Expected positions}
        \label{fig:Grid}
    \end{figure}

\section{Results}
\label{sec:Results}

    \subsection{Evaluation on the grid of sources - First test dataset}
    \label{sec:grid}

\begin{figure}[ht]
        \centering
        \includegraphics[scale = 0.35]{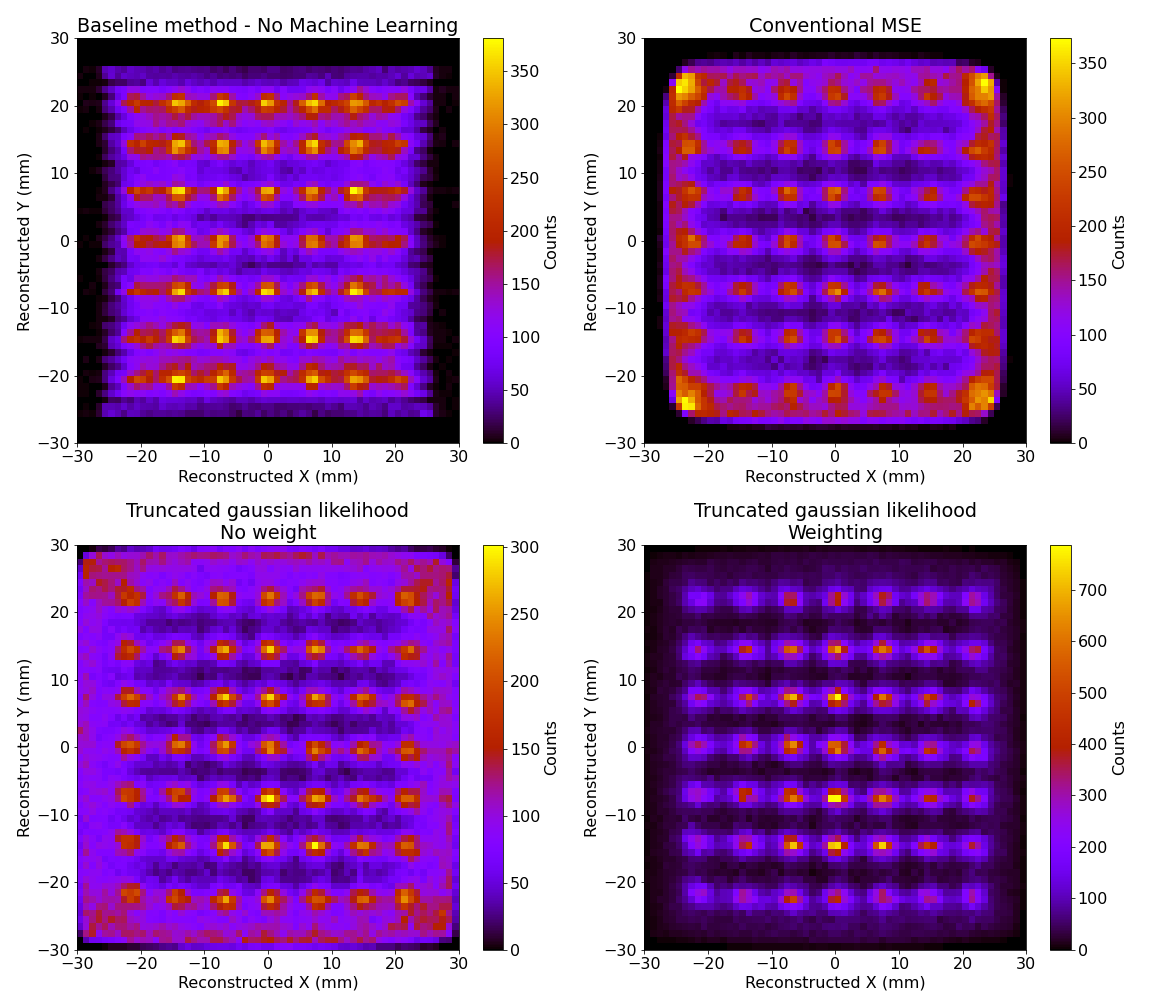}
        \caption{Grid reconstruction by the different methods}
        \label{fig:Resgrid}
    \end{figure}

    The positions of the grid of sources are represented on figure \ref{fig:Grid}. This configuration helps visualising directly some properties of the different reconstruction algorithms. The results are presented on figure \ref{fig:Resgrid}.

    They show that the algorithms are all able to reconstruct the sources in the center area of the detector, between -20 mm and +20 mm in both directions. However, the baseline method and the conventional Neural Network associated to a MSE loss function are not able to give a prediction on the edge of the detector. These predictions are even incorrectly attributed to other position at -25 mm and +25 mm for the conventional MSE approach. This "folding" of the predictions will imply a lower confidence of the reconstruction in this area, as we show in section \ref{sec:uniform}. On the contrary, the use of the truncated Gaussian likelihood is able to provide positions close to the edge and avoid the "folding" effect. This property shows a first advantage of the use of this custom loss function.

    The bottom right figure introduces the benefits of the uncertainty evaluation. To produce this figure, we use a weighting on each event $i$ according to the predicted uncertainties terms $\sigma_{\theta,x}(\boldsymbol{\mathrm{s_i}})$ for the X position, and $\sigma_{\theta,y}(\boldsymbol{\mathrm{s_i}})$ for the $Y$ position. The weights are computed using the following equation:

        \begin{equation}
        \label{eq:weighting}
            w_{i} = \frac{(\sigma_{\theta,x}(\boldsymbol{\mathrm{s_i}})^2 + \sigma_{\theta,y}(\boldsymbol{\mathrm{s_i}})^2)^{-1}}{F},
        \end{equation}

    where $F$ is a normalization factor such that $\sum_{i}(w_{i})$ is equal to the number of detected events on the considered bin of the histogram.

    \begin{equation}
        \label{eq:norm_factor}
            F = \sum_{i \in \mathrm{bin}}{(\sigma_{\theta,x}(\boldsymbol{\mathrm{s_i}})^2 + \sigma_{\theta,y}(\boldsymbol{\mathrm{s_i}})^2)^{-1}}
        \end{equation}
    
    This weighting penalizes the events with a high predicted uncertainty. The result shows a reduction of the spreading effect, making the reconstruction more accurate on the detector. We quantify this gain in accuracy in section \ref{sec:uniform}.
    
    \subsection{Evaluation on the uniform simulation - Second test dataset}
    \label{sec:uniform}

    \begin{figure}[h]
        \centering
        \includegraphics[scale = 0.35]{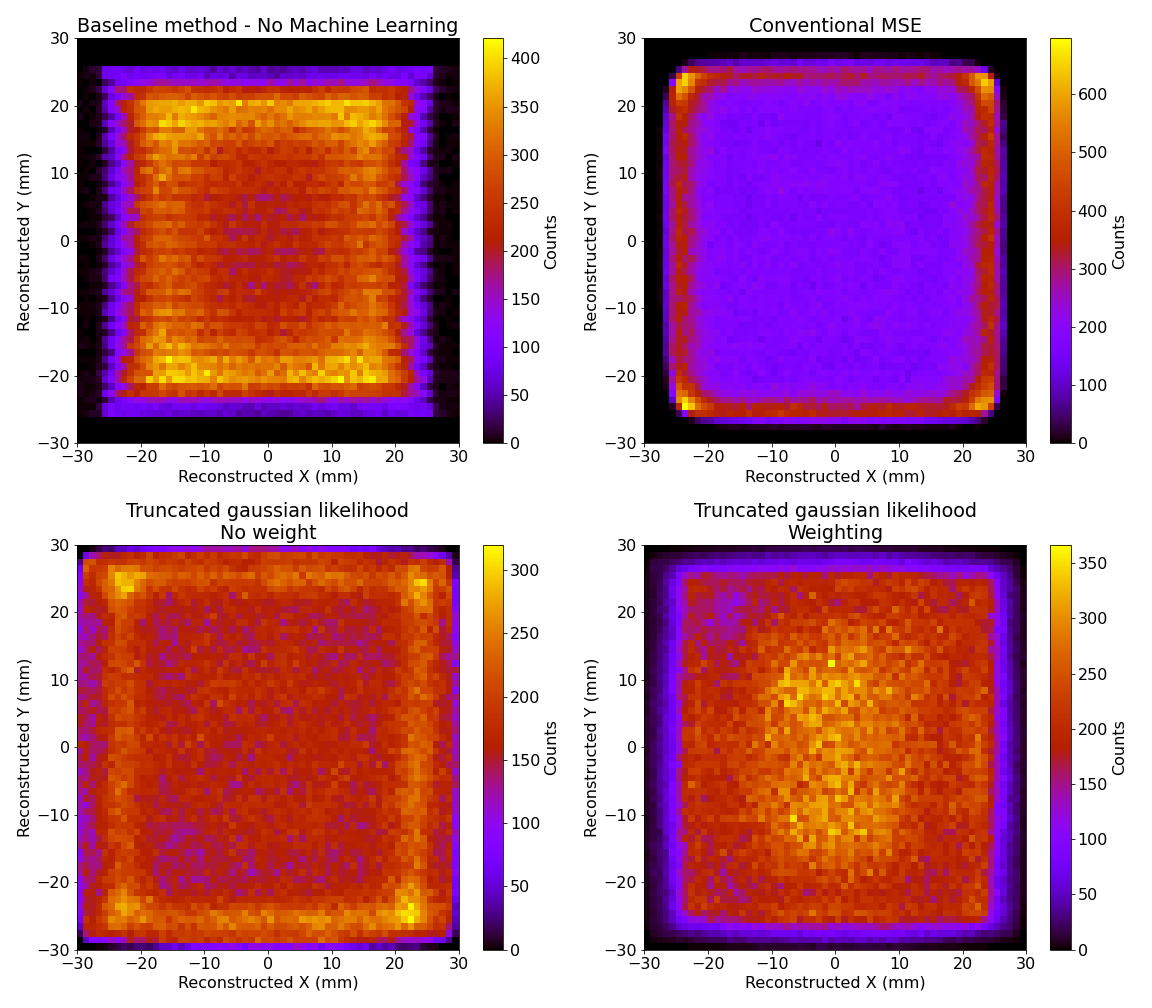}
        \caption{Reconstruction of the uniform simulation}
        \label{fig:ResUnif}
    \end{figure}

    For the global evaluation of the performances, we use the simulation of a uniform distribution of sources (cube source) in front of the detector. The direct reconstruction provided by the algorithms is shown on figure \ref{fig:ResUnif}. The results show the same properties as the grid simulation, especially the inability of the baseline method and the conventional MSE to reconstruct the edges of the detector. We can see a partial "folding" effect on the reconstruction with the truncated Gaussian likelihood. This effect is attenuated by applying the weighting as defined in equation \ref{eq:weighting}.

\begin{figure}[h!]
        \centering
        \includegraphics[scale = 0.35]{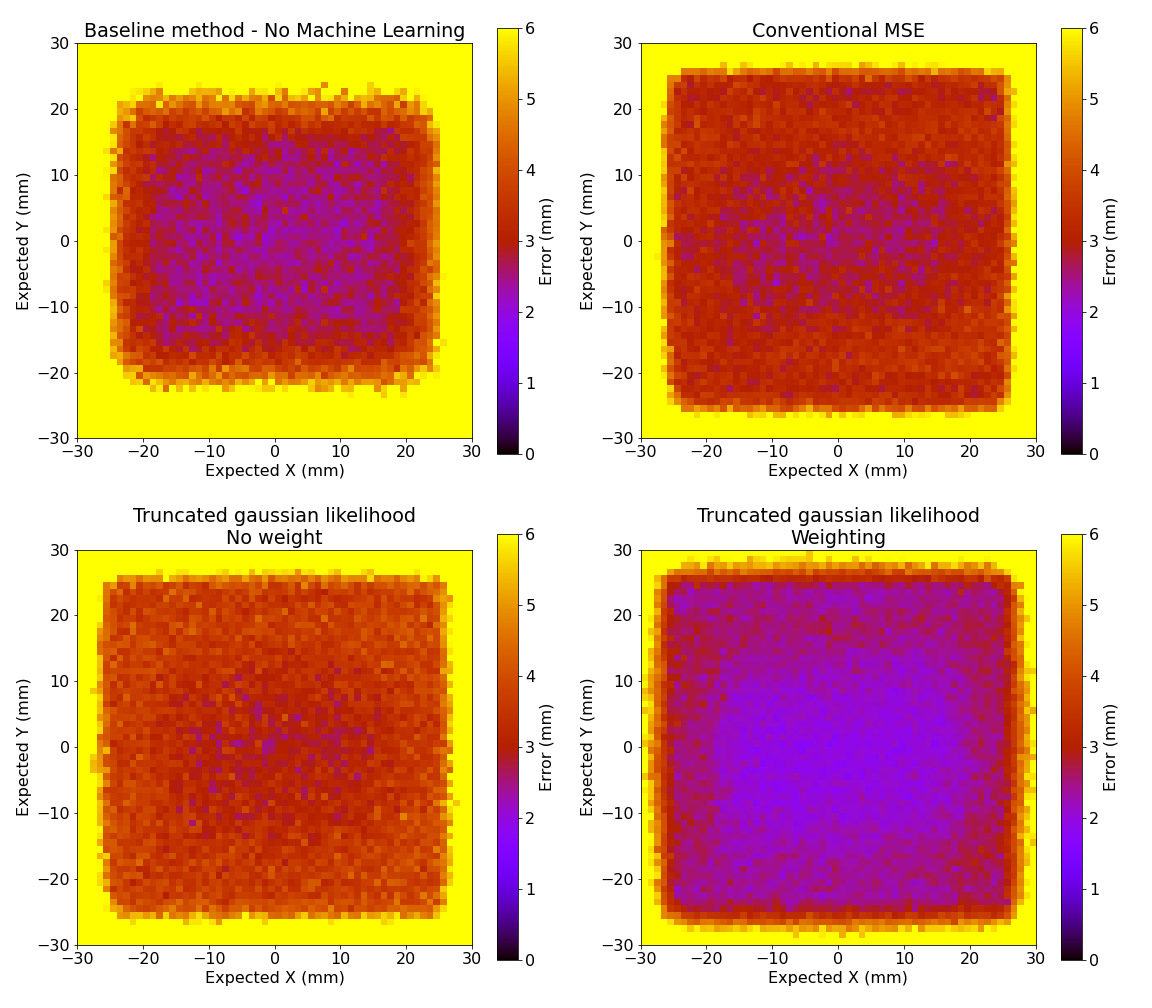}
        \caption{2D error reconstruction according to the true position, corresponding to the spread of the reconstruction - The color scale has been threshold at 6 mm for visualisation purpose}
        \label{fig:ResSpread}
    \end{figure}

    Figure \ref{fig:ResSpread} shows for each true position the average 2D distance to the reconstructed position. 
    A high average 2D distance in this case corresponds to a high spread of the reconstructions. 
    At the center of the detector, the baseline method provides less spread reconstructions, between 2 and 3 mm, than the conventional MSE and our approach with the truncated Gaussian likelihood, between 3 and 4 mm. However, the area of high spread, over 6 mm, is larger for the baseline method than the machine learning method. 
    We show later that this default highly affects the precision of the reconstruction, even at the center of the detector.

    The use of the weighting helps the truncated Gaussian likelihood to give more importance to the most certain events, leading to a spread between 1 and 2 mm at the center of the detector, between 2 and 4 mm in the intermediate area. The area with the higher spread, over 6 mm, is thiner than the other algorithms.    

\begin{figure}[h!]
        \centering
        \includegraphics[scale = 0.35]{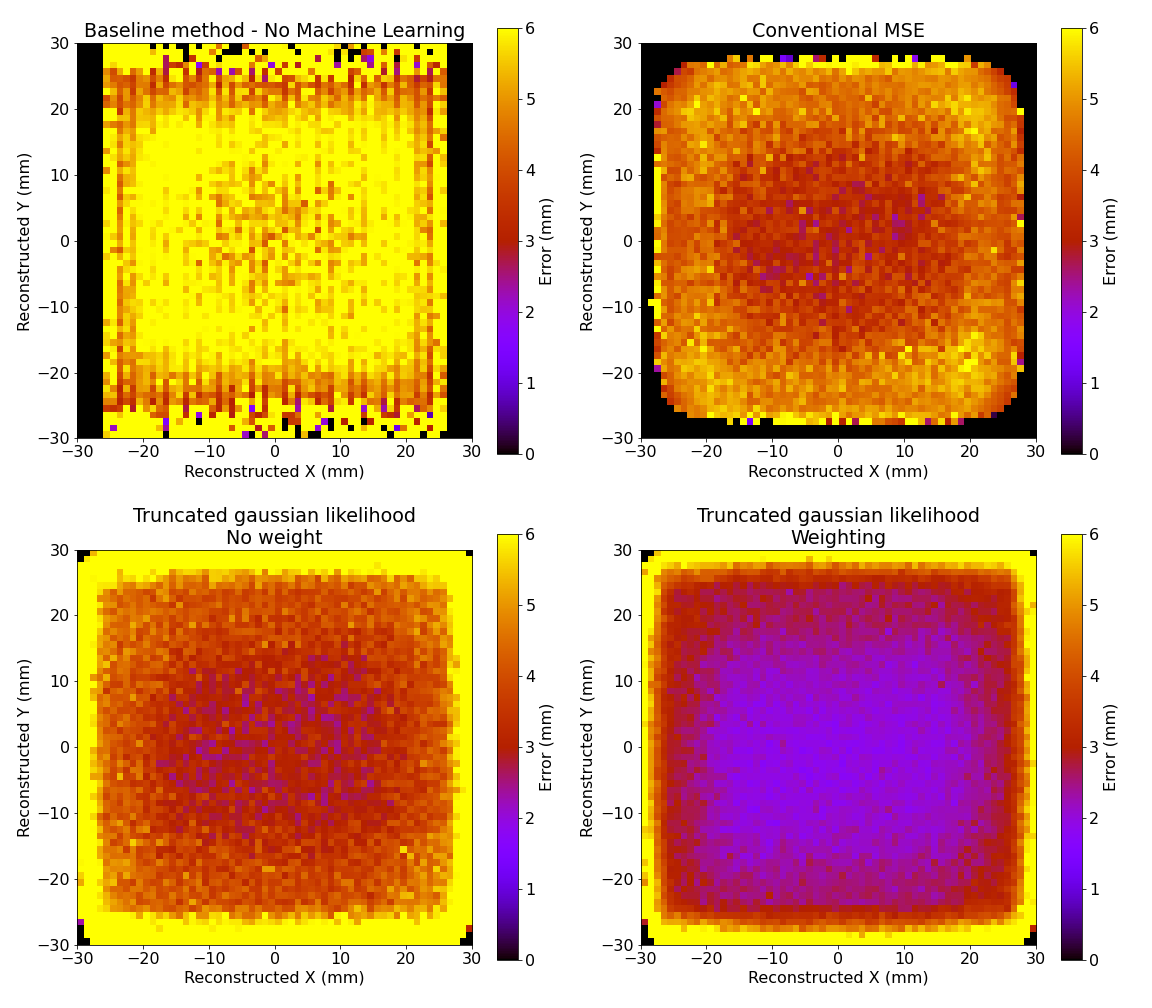}
        \caption{2D error reconstruction according to the predicted position, corresponding to the precision of the reconstruction - The color scale has been threshold at 6 mm for visualisation purpose}
        \label{fig:ResPrecision}
    \end{figure}

    Figure \ref{fig:ResPrecision} shows for each predicted position the average 2D distance to the true position. 
    A high average 2D distance in this case corresponds to a low precision of the reconstructions. 
    Black areas correspond to area with no predicted position due to the inability of the algorithms to reconstruct some positions, as shown on figure \ref{fig:ResUnif}.
    The baseline method gives very low accurate results, mostly over 6 mm, even at the center of the detector. 
    By combining with the previous information, we can give the following interpretation:
            \begin{itemize}
            \item
            if the algorithm provides a localisation at the center of the detector, we cannot trust the algorithm since the reconstruction error is over 5 mm, as shown on figure \ref{fig:ResPrecision} ;
            \item 
            if the true position is the center of the detector, we can trust the algorithm because the reconstruction error is between 2 and 3 mm, as shown on figure \ref{fig:ResSpread}. However, for a real application, we cannot access to the true position, then we cannot exploit this information.
        \end{itemize}

    The conventional MSE shows a better precision than the baseline method, with a reconstruction error between 3 mm at the center of the detector to 5 mm close to the edges.
    The areas with very high reconstruction error, over 6 mm, are limited. 
    The truncated Gaussian likelihood shows similar performances as the conventional MSE for the precision, with the ability to reconstruct the edges, with less precision. 
    The application of the weighting improves the results, providing a precision between 1 and 2 mm at the center of the detector, between 3 and 5 mm in the intermediate area and over 6 mm in a thin band close to the edge. 
    This band corresponds to the part of the detector without optical layer and it is expected to have a degradation of the performances in this area.

\begin{figure}[h!]
        \center
        \includegraphics[scale = 0.35]{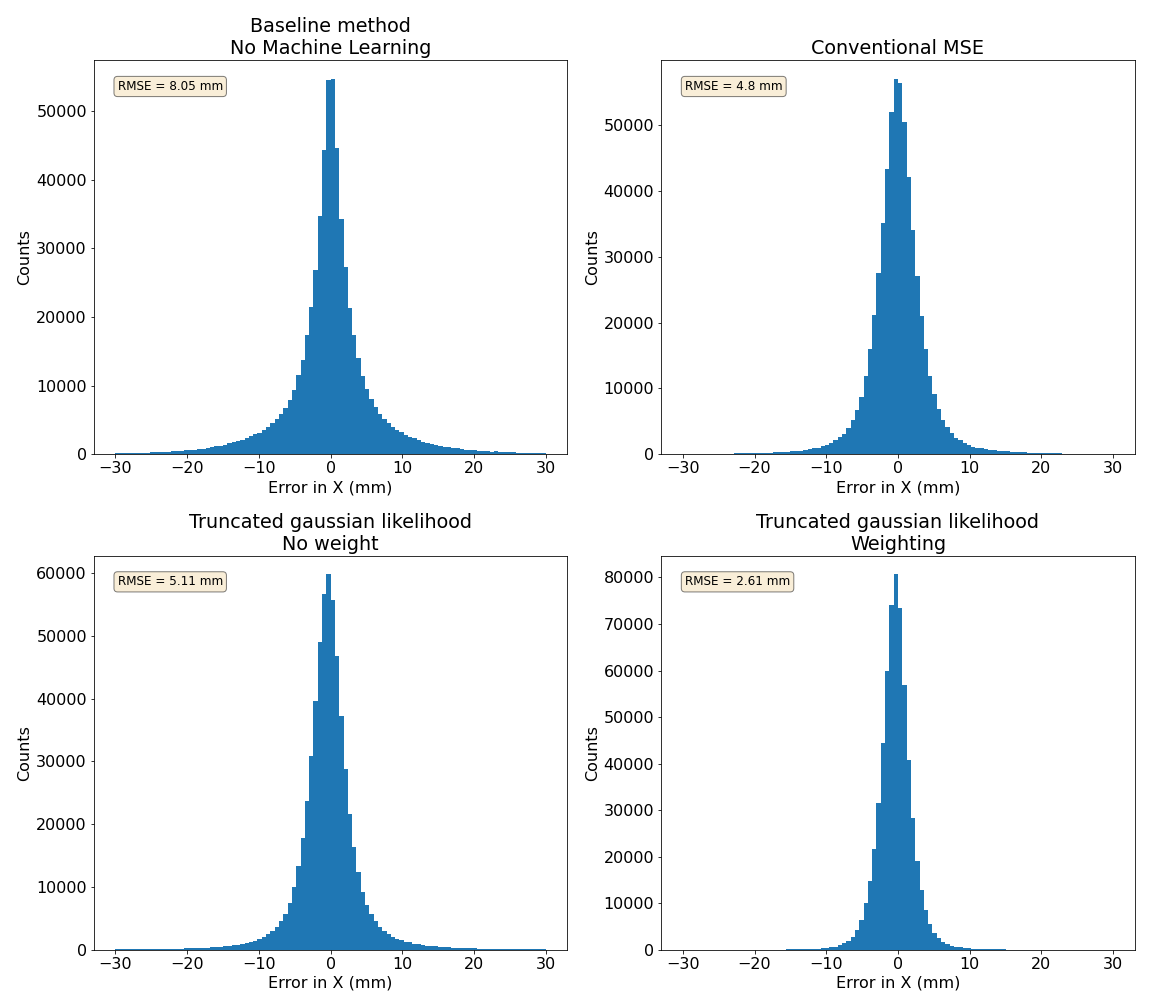}
        \caption{Global histogram of reconstruction errors on the X position}
        \label{fig:HistoX}
    \end{figure}

\begin{figure}[h!]
        \centering
        \includegraphics[scale = 0.35]{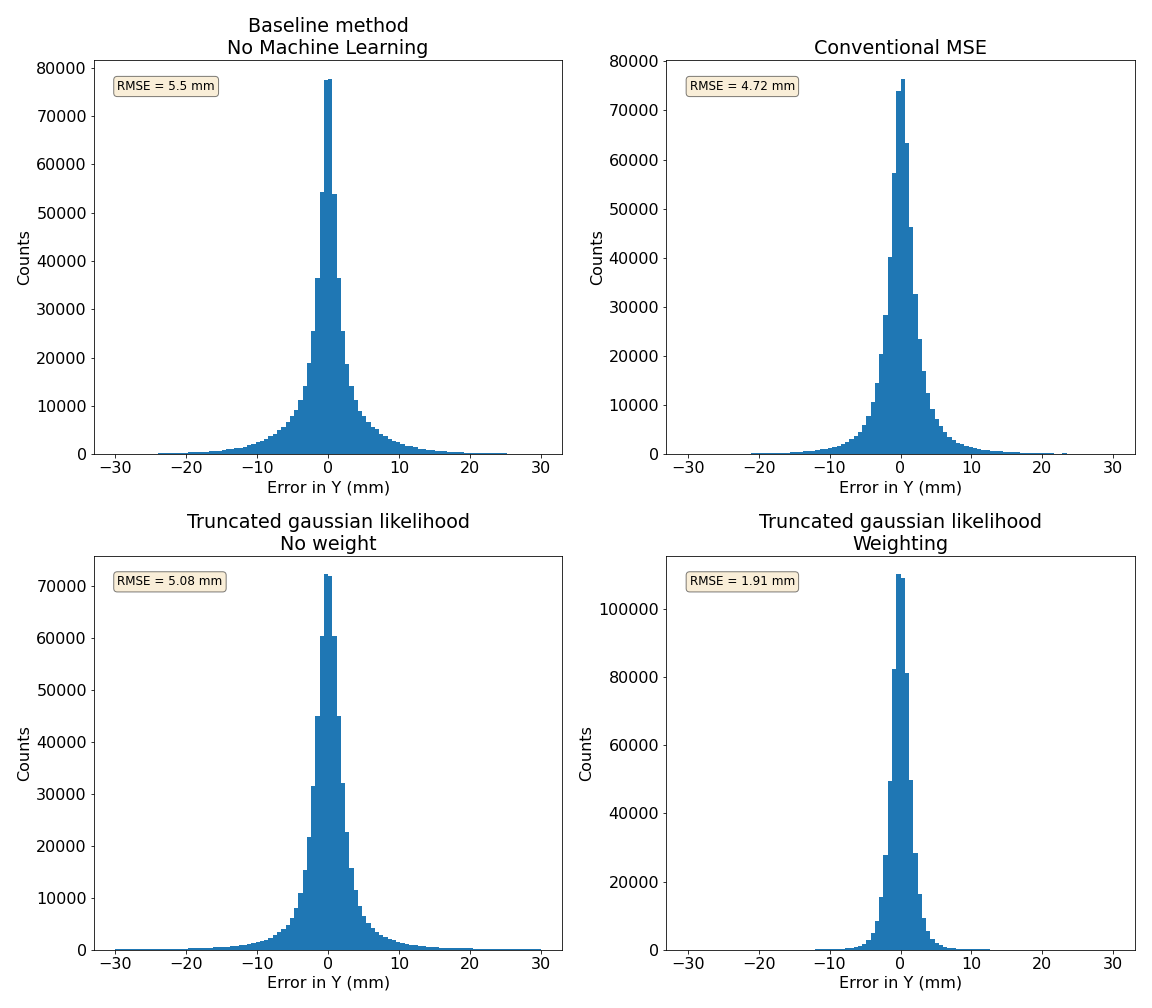}
        \caption{Global histogram of reconstruction errors on the Y position}
        \label{fig:HistoY}
    \end{figure}

    Finally, figures \ref{fig:HistoX} and \ref{fig:HistoY} show the overall results, considering the whole test dataset. The histograms represent the global reconstruction errors, without selecting any position. We also compute the Root Mean Squared Error (RMSE) of the predictions. 
    
    For both coordinates, the baseline method shows the worse performances. This result shows the advantage to use a Machine Learning approach to address this reconstruction problem, due to the complex link between the observable variables and the positions.

    We apply a new 1D-weighting dedicated to each coordinate: 

        \begin{equation}
        \label{eq:weightingX}
            w_{X,i} = \frac{(\sigma_{\theta,x}(\boldsymbol{\mathrm{s_i}}))^{-2}}{\sum_{j}{(\sigma_{\theta,x}(\boldsymbol{\mathrm{s_j}}))^{-2}}}
        \end{equation}
        \begin{equation}
        \label{eq:weightingY}
            w_{Y,i} = \frac{(\sigma_{\theta,y}(\boldsymbol{\mathrm{s_i}}))^{-2}}{\sum_{j}{(\sigma_{\theta,y}(\boldsymbol{\mathrm{s_j}}))^{-2}}}
        \end{equation}

    On average, the conventional MSE gives slightly better results than the truncated Gaussian likelihood in terms of RMSE. We can expect this behaviour because the MSE approach is specifically designed to minimize the MSE, thus its squared root the RMSE.
    
    Table \ref{tab:tableperfo} summarizes several performances such as the mean error (the bias) on the histogram, RMSE and the Standard Deviation (spread of the distribution without bias). A small bias is observed for the reconstructions with the truncated Gaussian likelihood methods for the X coordinate, under 1 mm. On the contrary, the conventional MSE presents a bias for the Y coordinate reconstruction. On the other metrics, the weighting applied to the truncated Gaussian likelihood outperforms the other methods: the application of this weighting helps improving the performances by giving less importance to uncertain reconstructions. Especially, the RMSE reaches 2.61 mm for the X coordinate and 1.91 mm for the Y coordinate.
    
    \renewcommand\tabularxcolumn[1]{m{#1}}
    \begin{table}[h!]
    \label{tab:tableperfo}
    \begin{center}
    \begin{tabularx}{0.8\textwidth} { 
    | >{\centering\arraybackslash}X 
    | >{\centering\arraybackslash}X 
    | >{\centering\arraybackslash}X 
    | >{\centering\arraybackslash}X 
    | >{\centering\arraybackslash}X | }
     \hline
     Quantity & Baseline method & Conventional MSE & Truncated Gaussian lik. & Truncated Gaussian lik. weighting \\ [0.5ex] 
     \hline\hline
     Mean error X (mm) & \textbf{-0.01} & \textbf{0.01} & -0.32 & -0.39 \\ 
     \hline
     RMSE X (mm) & 8.05 & 4.8 & 5.11 & \textbf{2.61} \\
     \hline
     Std dev X (mm) & 8.05 & 4.8 & 5.1 & \textbf{2.58} \\
     \hline\hline
     Mean error Y (mm) & \textbf{0.00} & 0.19 & 0.04 & 0.02 \\ 
     \hline
     RMSE Y (mm) & 5.5 & 4.72 & 5.08 & \textbf{1.91} \\
     \hline
     Std dev Y (mm) & 5.5 & 4.72 & 5.08 & \textbf{1.91} \\
     \hline
    \end{tabularx}
    \caption{\label{tab:tableperfo}Table of performances. Best values are in bold.}
    \end{center}
    \end{table}

    \subsection{Calibration of the uncertainties}
        \label{sec:uncertaintyperf}

    The evaluation of the performances shows the advantage of using the uncertainty prediction as a weighting of the uncertain reconstruction. However, these results do not provide accurate elements to assess the calibration of the uncertainties. In this section, we provide an evaluation of the quality of the uncertainty estimation by using a coverage plot.
    
    To produce this plot, we consider a probability level $\alpha$. For each event $i$, our neural network provides a prediction of the parameters $\mu^{(i)}_\theta$ and $\sigma^{(i)}_\theta$ for a truncated Gaussian distribution, with bounds $a$ and $b$. Under the hypothesis of this distribution, we compute the Prediction Intervals (PI) at level $\alpha$ for each event $I^{(i)}(\alpha)$ such that:

        \begin{equation}
        \label{eq:predictiveproba}
            p(y_{\mathrm{true}} \in I^{(i)}(\alpha)) = \alpha
        \end{equation}

    where $y_{\mathrm{true}}$ is the expected value. There is an infinite number of possible intervals that correspond to this condition, we choose the interval that is centered on the expected median.
    Then, we compute the number of events whose expected value belongs indeed to the Prediction Interval, which leads to the Prediction Interval Coverage Probability (PICP) at level $\alpha$:
    
        \begin{equation}
        \label{eq:predictiveproba1}
            \mathrm{PICP}(\alpha) = \frac{|\{y_{\mathrm{true}} \in I^{(i)}(\alpha)\}_i|}{N}
        \end{equation}
    
    where $|E|$ is the cardinal of the ensemble $E$ and $N$ is the total number of events. The PICP can be seen as the empirical frequency of the true values that indeed belong to the Prediction Interval.
    Finally, we compare the $\mathrm{PICP}(\alpha)$ to the probability level $\alpha$:
            \begin{itemize}
            \item
            if $\mathrm{PICP}(\alpha) = \alpha$, we have a perfect calibration;
            \item 
            if $\mathrm{PICP}(\alpha) > \alpha$, the model is under-confident, or in other word, conservative;
            \item 
            if $\mathrm{PICP}(\alpha) < \alpha$, the model is over-confident.
        \end{itemize}

\begin{figure}[ht]
        \centering
        \includegraphics[width=1.\textwidth]{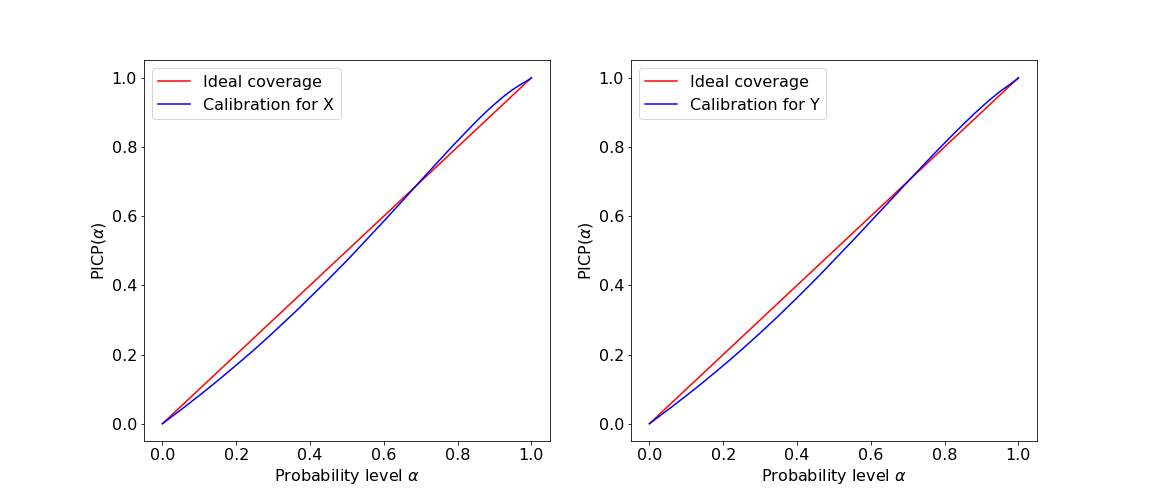}
        \caption{Coverage plots - The models are slightly too confident for probability levels lower than 75 \% and slightly too conservative for probability levels higher than 75 \%.}
        \label{fig:Coverage}
    \end{figure}

    Figure \ref{fig:Coverage} shows the results for different probability levels $\alpha$. For both coordinates, the PICP is close to the value $\alpha$ which means the Prediction Interval can be trusted in average. The model is slightly over-confident for probability levels under 75 \% and conservative for higher probability levels. For instance, Prediction Intervals at 95 \% are conservative.
    
\section{Discussion}
\label{sec:Discussion}

\subsection{Benefits of the study for the detector reconstruction and the PET imaging}

The use of Machine Learning approaches to reconstruct gamma photon position of interaction in monolithic crystals is a hot topic nowadays. 
For PET applications, most of the time it is a LYSO scintillator of several tenth of millimeters size and about 1~cm thickness coupled with pixelated silicon photomultiplier (SiPM).
For example, in~\cite{belov_resolution_2023} the spatial resolution obtained with a feed-forward neural network is 0.74 $\pm$ 0.01~mm for the XY plane and an average 1.01 $\pm$ 0.01~mm for the Z coordinate. 
Similar results were obtained in~\cite{freire_performance_2022} with one multilayer perceptron with five hidden layers and 100 nodes. In~\cite{carra_neural_2022}, the sub-millimeter precision in XY plane is obtained with an algorithm based on neural networks integrated into a second neural network for simultaneous estimation of the event position and timestamp. 
Even a preclinical PET system with an annular monolithic LYSO scintillator was proposed by~\cite{jaliparthi_deep_2021} with an inner diameter of 6~cm with minimum transaxial scintillator thickness of about 1~cm covered by SiPMs. 
Authors of these study obtained the spatial resolution on reconstructed interaction position of about half of millimeter with a ten-layer deep residual-convolutional neural networks. 
Sub-millimeter precise photon interaction position determination in CeBr$_{3}$ crystal and LaBr3$_{3}$:Ce~\cite{kawula_sub-millimeter_2021} for Compton Camera imaging system was obtained with a newly designed convolutional neural network of five layers.

As shown in these recent studies and in section \ref{sec:Results}, the Machine Learning approaches provide promising and more accurate reconstructions than the baseline methods, only based on the knowledge of the detector. 
In contrast to prior research with LYSO or other scintillation only crystals, this current study aims a precise reconstruction of the gamma photon
interaction in a \PWO crystal with  faster (Cherenkov effect) but fewer (lower light yield) optical photons. 
Moreover, in contrary to pixelated SiPM with one readout channel per pixel, a MCP-PMT with fewer readout channels ($2 \times 32$ instead of $32 \times 32$) is used. 
These differences introduce additional complexity on the precise position reconstruction.
The coordinate along the transmission lines (X-coordinate) turned out to be complex to reconstruct, but our machine learning models were able to learn to address this complexity.

For monolithic crystals, the so-called edge-effect worsens the spatial localization performance toward the detector borders~\cite{freire_performance_2022}. 
In our study, the use of a dedicated loss function, the truncated Gaussian likelihood, helps recovering some dead area at the edges that are not reconstructed by the baseline methods or the conventional MSE. 
This effect can be explained by the fact that no interaction can be observed outside of the detector, so the MSE avoids to provide predictions close to the area outside of the detector and consequently, close to the edges. 
Because we include this prior knowledge in the truncated Gaussian likelihood, the model trained with this loss function can predict positions close to the edges.

Different from previous research, an uncertainty on the gamma photon interaction reconstruction is provided in this study by the Density Neural Network.
This additional information is included by a weighting in the performance evaluation. 
We have shown that it significantly improves the performances. 
We can object that this improvement is "artificial" because we do not improve the prediction, we give less importance in the metrics to the uncertain prediction. 
However, it is very promising for the application in PET imaging because the tomographic image reconstruction relies on the processing of a large number of events. 
If some events carry less information, they can degrade the SNR in the reconstructed image. 
As a consequence, accounting for the uncertainty information in the tomographic reconstruction shall increase the SNR in the image.
The uncertainty can be included in a spatial resolution model adapted to each individual event,
enabling an event-based resolution modeling image reconstruction. 
The proposed approach would be innovative for PET imaging, with the best of our knowledge, and its impact on the reconstructed image quality must be assessed.
The combination of the proposed gamma photon interaction reconstruction in the detector with the full PET image reconstruction is planned for future works.

Finally, the calibration of the uncertainties show that their prediction is satisfying, although not perfect. The disagreement between the probability level and the empirical frequency can come from two factors:
            \begin{itemize}
            \item
            the prediction of the neural networks can be improved by optimizing the model and the training phase;
            \item 
            the truncated Gaussian hypothesis could not be the best representation of the uncertainty. Further works are planned to use other probability distribution functions, under the constraint that these probability distributions must not diverge and be differentiable so that they can be used as loss functions.
            \end{itemize}

\subsection{Generalization of the methodology}

Through this paper, we want to highlight the use of a specific methodology, which can be beneficial for many applications of Deep Neural Network in regression tasks. This approach relies on the paradigm of uncertainty estimation, which is a growing topic in the scientific literature on Artificial Intelligence. Several approaches are usually proposed, such as Bayesian Neural Network or Deep Ensembles \cite{gawlikowski2021}. The use of the Density Neural Network is an interesting approach when the uncertainties are dominated by fluctuations and random phenomenon in the data, such as the source of variability during the scintillation process in our detectors. This approach brings additional information thanks to the uncertainties that can be used to gain information on the prediction of neural networks and can be exploited in numerous applications. A dedicated methodology of exploitation and validation has been presented in this paper.

More generally, we can describe the methodology as follows:
\begin{itemize}
            \item
            The design of a dedicated loss function, based on a negative log-likelihood of an assumed distribution. This distribution is often assumed as a Gaussian distribution, but it is not restricted to this type of distribution and can be adapted to the problem to address. In our case, we decided to add physical constraints through a truncation of a Gaussian distribution according to the detector edges, which requires a specific implementation of the loss function. We show that this approach brings improvements for reconstructing specific events, especially close to the detector.
            \item 
            The estimated uncertainties can be exploited to complete the prediction of the neural network. The uncertainty can be evaluated event by event to assess the confidence on the prediction of the model. In our case, we show that we can statistically improve the reconstruction resolution and this feature will be helpful in the exploitation of the detector. Uncertain events are consequently considered as noisy and can be discriminated through this method. Applications applying Neural Networks to analyse signals or sensors data would draw benefits from this approach.
            \item
            It must be kept in mind that the uncertainties are self-estimated by the Neural Network, thus they are prone to the same biases as the classical prediction. Consequently, the validation of the estimated uncertainty is required and the use of coverage plot is a possible approach to empirically verify the relevance and reliability of these uncertainties.
            \end{itemize}

\section{Conclusion}
\label{sec:Conclusion}

Our objective to predict the 2D position of gamma interaction in the ClearMind detector is successfully achieved thanks to Machine Learning methods. Our neural network outperforms the baseline approaches based on physical knowledge only and used as inputs of the model. The introduction of Density Neural Networks provides a reliable estimation of the uncertainty in the prediction of the neural network that can be used to discriminate reconstructions that are likely less accurate. We aim to draw benefits from this method in further works for PET image reconstruction. Moreover, we exploit the flexibility of the Density Neural Network approach to design a truncated gaussian loss function based on physical constraints. This property helps for the reconstruction of events in areas far from the center of the detector, increasing the exploitable surface of the crystal. We want to highlight that this methodology is generic and can be adapted to other use cases, as far as we are able to introduce such constraints in the loss function.

The preprocessing step is also important to reduce the number of input variables, thus to get a compact neural network that can be embedded as close as possible to the detector. The computation time on test data on a conventional CPU is 14 s for the 600 000 events, corresponding to 23 µs per event.

In outlooks, we want to conduct a sensitivity analysis on the input variables. The objective is to study the importance of the preprocessed observables for the prediction of the positions. We will verify if we find the expected correlation between the prediction of the positions and the identified observables given by the physical expertise. Moreover, the analysis will be performed on the uncertainty estimation in order to understand which quantities have the highest influence on the uncertainty in the reconstruction. We will compare the performances in reconstruction with the use of the raw data, in order to assess the possible loss of information due to this preprocessing. In further works, we will also apply this methodology for the Depth of Interaction, energy and time reconstructions in order to obtain complete information on the gamma ray reconstructions. Moreover, these results will be exploited to complete PET image reconstruction in dedicated studies, in order to improve the SNR on the image by the weighting or the discrimination of uncertain reconstruction.

\section*{Acknowledgments}

This project has received funding from the European Union's Horizon 2020 research and innovation program under the Marie Sk{\l}odowska-Curie grant agreement No 800 945. 
We are grateful for the support and seed funding from the CEA, 
Programme Exploratoire Bottom-Up, under grant No. 17P103-CLEAR-MIND, the Cross-Disciplinary Program on Numerical Simulation of CEA for the financial support of project AAIMME,
and the French National Research Agency under grant No. ANR-19-CE19-0009-01.

\section{Appendix: Statistical Processing of raw Waveforms}
\label{sec:Prepro}

The following paragraphs describes the preprocessed observables we developed for the ClearMind detector in addition to the description in section \ref{sec:WaveProcess}.
First, for each line that triggered the SAMPIC acquisition, let's define $F_{l,\mathrm{Left}}(t_j)$ and $F_{l,\mathrm{Right}}(t_j)$, the digitized pulse shapes (Left and Right of transmission line) at time $t_j$. \newline
Then, before computing event observables, we calculate for each trigered line:
 \begin{itemize}
    \item
    The times of the first samples of the pulse shapes $T_{0,\mathrm{Left}}$, $T_{0,\mathrm{Right}}$
    \item
    The times calculated on the rising edges of the pulses, extrapolated to the half height between the sampled values, $T_{l,\mathrm{Left}}$ and $T_{l,\mathrm{Right}}$.
    \item
    The time lapse when the pulses have exceeded the WaveCatcher trigger threshold.
    \item
    The time lapse when the pulses have saturated the WaveCatcher.
    \item
    The sum of the charges collected at both ends of the transmission lines $C_{l}$.
\end{itemize}

The selected observables can be classified into several categories.\newline
First, a general parameter stores the number of transmission lines that have acquired a signal.\newline

Then we store the observables expected to be correlated to the \emph{interaction time} of the gamma photon in the crystal. These are :
\begin{itemize}
    \item
    The time of the first SAMPIC sample recorded on the first transmission line that acquired a signal. This time $T_{\mathrm{First}}$ is then subtracted from all the time variables associated with the event.
    \item
    The time of the first photoelectron detected on the event. We store the lowest photoelectron time value (defined as  $0.5 \times (T_{l,\mathrm{Left}}$ + $T_{l,\mathrm{Right}}$)) of all the lines that triggered the acquisition.
\end{itemize}

We add four bias observables, intended to help the neural network to decorrelate possible stacking effects on the pulse shapes. These are
\begin{itemize}
    \item
     the time lapse beyond 50\% of the amplitude on the pulse shapes $F_{l,\mathrm{Left}}(t_j)$) and $F_{l,\mathrm{Right}}(t_j)$ (one observable for Left pulse shape and one for Right)
    \item
     the computed pulses' rise time (one observable per side)
\end{itemize}

Then we store the observables expected to be correlated to the \emph{energy deposited} by the interaction in the detector which is a quantity that is planned to be reconstructed for future works. These observables are also expected to bring information to estimate the uncertainties on the interaction position. We store : 
\begin{itemize}
    \item
    The sum of the charges collected over all lines $C_{T}$.     
    \item
    The sum of the time lapse when pulses have exceeded the triggering threshold of the WaveCatcher $T_{\mathrm{Thres}}$ over all lines.
    \item
    The sum over of the time lapse when pulses have saturated the WaveCatcher (bias observable), $T_{\mathrm{Sat}}$.
\end{itemize}  
Then we store observables correlated to the position on the axis perpendicular to the transmission lines $x_{\mathrm{pos}}$ and along transmission line $y_{\mathrm{pos}}$, as explained in section \ref{sec:WaveProcess}. \newline

Finally, we compute quantities expected to be correlated to the \emph{depth of interaction} (DOI) of the gamma in the crystal, used for future works to achieve 3D reconstruction. We use the fact that the farther the gamma interaction occurs from the photoelectric layer, the more photoelectrons are produced over a large area. The depth of interaction is thus correlated to the dispersion of the produced photoelectrons. We can also expect that this dispersion is relevant for our neural network to estimate the uncertainty on $x_{\mathrm{pos}}$ and $y_{\mathrm{pos}}$. To quantify this dispersion we first calculate the distribution: $ \lvert l-\mathrm{Med}_{L}\rvert $ , weighted by the charge measured on the line l, $C_{l}$.
$\mathrm{Med}_{l} = \mathrm{median}( \lvert l-Med_{L}\rvert , C_{l})$.
The two first observables are then :
 \begin{itemize}
    \item
    the median($\lvert l-\mathrm{Med}_{L} \rvert, C_{l}$) and
    \item
    the mean($\lvert l-\mathrm{Mean}_{L} \rvert, C_{l})$
\end{itemize}

To quantify the dispersion along the transmission lines, we use the same algorithms, but on the previously computed $\Delta t_{l,p}$ and ${\mathrm{Bipol}}_{l,p}$ distributions (section \ref{sec:WaveProcess}), weighted by the charge collected in the pulses. The observables are then : 
 \begin{itemize}
    \item
    the median($\lvert \Delta t_{l,p}-\mathrm{Med}_{\Delta t} \rvert, C_{l,p}$), 
    \item
    the mean($\lvert \Delta t_{l,p}-\mathrm{Mean}_{\Delta t} \rvert, C_{l,p}$),
    \item
    the median($\lvert \mathrm{Bipol}_{l,p}-\mathrm{Med}_{\mathrm{Bipol}} \rvert, C_{l,p}$) 
    \item
    the mean($\lvert \mathrm{Bipol}_{l,p}-\mathrm{Mean}_{\mathrm{Bipol}} \rvert, C_{l,p}$) 
  \end{itemize}
    
We did not devise until now any bias observables on the depth of interaction parameter.

    \section*{Appendix: Median algorithm on weighted distributions}
    
\label{sec:MedianAlgo}
Following is our algorithm for computing the median of a binned distribution, written in C++. It uses the STL library to sort the input values. Instead of returning the central value of the bin where the mid-point value of the distribution happen, we decided to make use of the values of the neighbouring bins to qualify the result.

\begin{verbatim}
bool CompPairValues(const PairValues& Pair1, const PairValues& Pair2)
{ return (Pair1.Val<Pair2.Val); }

// Compute the mid point of a set of values VValue of weigths Vweight
double Calc_Median_Weighted( vector<double> Vvalues, vector<double> Vweigth, bool PrintInside) {
  unsigned long NbPair = Vvalues.size();
  if(NbPair==0) { cerr<<"Error in DY_Calc_Median_Weighted: Vvalues empty" <<endl;
    exit(-1);
}

// We store values in pairs
  vector<PairValues> VPair(NbPair);
  double Half_SomWeight = 0.;
  for( unsigned long ind= 0; ind < NbPair; ind++) {
    VPair[ind].Val = Vvalues[ind];
    VPair[ind].weigth = Vweigth[ind];
    Half_SomWeight += VPair[ind].weigth;
  }
  Half_SomWeight *= 0.5;

  // We sort values, stable_sort ou sort
  std::stable_sort( VPair.begin(), VPair.end(), CompPairValues);

 // We find mid-weight point
  double ValPlus = 0.;  double ValMinus = 0.;
  double SomWeigPlus = 0.; double SomWeigMinus = 0.;
  unsigned long IndexPlus =0;
  for( unsigned long Index =0; Index<NbPair; Index++)
  { SomWeigPlus+=  VPair[Index].weigth; 
    ValPlus= VPair[Index].Val; 
    IndexPlus= Index;
    if(SomWeigPlus>Half_SomWeight) break;
    SomWeigMinus = SomWeigPlus; 
    ValMinus = ValPlus;
  }

  // Compute the Median 
  double ValueMedian, ValueMedian2, Median;
  double SomWeigMinDes ;
  if( (IndexPlus == 0) || (IndexPlus == (NbPair-1)) ) 
    Median = ValPlus;
  else {
    double Un_WeightIndPlus= 1./VPair[IndexPlus].weigth;
    // Rising Option 
    double DiffSom = Half_SomWeight-SomWeigMinus;
    double DeltaVal = ValPlus-ValMinus;
    ValueMedian = ValMinus + DeltaVal*DiffSom*Un_WeightIndPlus;
    // Descending option
    double ValMinDes = VPair[IndexPlus+1].Val;
    SomWeigMinDes = Half_SomWeight*2. - SomWeigPlus;
    double DiffSom2 = Half_SomWeight - SomWeigMinDes;
    double DeltaVal2 = ValPlus-ValMinDes;
    ValueMedian2 = ValMinDes + DeltaVal2*DiffSom2*Un_WeightIndPlus;
    Median = 0.5*(ValueMedian+ValueMedian2);
  }
  if( PrintInside) // Print everything for debugging { } 
  return Median;
}
\end{verbatim}

\bibliographystyle{plain}
\bibliography{biblio/biblio,biblio/Detectors,biblio/Scintillators,biblio/PET}

\end{document}